\documentclass[a4paper]{article}

\title{Stochastic System with Colored Noise and Absorbing States:\\ Path Integral Solution}

\author{D.~O.~Kharchenko\footnote{dikh@ssu.sumy.ua}\\
Sumy State University, 40007 Sumy, Ukraine}
\date{\ }
\begin{document}
\maketitle

\begin{abstract}
The behavior of the most probable values of the order parameter $x$ and the
amplitude $\phi$ of conjugate force fluctuations is studied for a stochastic
system with a colored multiplicative noise with absorbing states. The phase
diagrams introduced as dependencies the noise self-correlation time vs
temperature and noise growth velocity are defined. It is shown that phase
half-plane $(x,\phi)$ can be split into isolated domains of large,
intermediate, and small values of $x$. System behavior in these domains is
studied by the probability represented as path integral. In the region $x\ll
1$, the trajectories converge to the point $x = \phi = 0$ for $0 < a < 1/2$ and
to $x = 0$, $\phi\to\infty$ for $1/2 < a \le 1$. In the former case, the
probability of realization of trajectories is finite, while in the latter case
it is vanishingly small, and an absorbing state can be formed.
\end{abstract}

A wide class of stochastic systems with multiplicative noise whose intensity is
a function of the stochastic variable $x$ determining the state of the
subsystem can exhibit absorbing states. An absorbing configuration is one in
which a system can get trapped, from which it can not escape \cite{Dickman}.
The most popular example of this kind is the Verch\"ulst model proposed for
explaining the population kinetics \cite{Horsthemke}; the analysis of this
model remains a topical question \cite{Geraschenko}. Another example is the
percolation model \cite{Munoz} in which a linear dependence on $x$ is observed
for noise intensity. Indicated models can be generalized by representing the
noise intensity as a stochastic variable power dependence with an arbitrary
exponent. A kinetic features of stochastic system with that kind noise was
studied in \cite{ftt} in the white noise approximation and with accounting
inhomogeneity of the space. An evolution of averages in that model was
discussed in \cite{myEprint}. A distinguishing feature of our approach is that
we shall study the dependencies of most probable values vs system parameters
using supposition of the noise correlation. This is realized on the basis of a
field scheme in which the probability functional has an exponential form with
the exponent that can be reduced (correct to the sign) to the standard action
\cite{ZJ}. Consequently, a description of phase transition in terms of the most
probable values corresponds to the application of the least action principle.
Such an approach makes it possible to extend the self-consistent field method
to a description of systems with multiplicative noise. This reveals the
following nontrivial peculiarity: the most probable values can tend either to
free energy minima or to points corresponding to unstable thermodynamic states
depending on the system parameters. The singular nature of noise indicating the
tendency of its intensity to zero for the value $x = 0$ of the order parameter
(more probably magnitude of $x$) is responsible for the emergence of a region
on the phase portrait in which all trajectories converge to the axis $x = 0$
over a finite time. In other words, an absorbing state is formed, in which the
system becomes closed \cite{Dickman}.

We consider following model. The time dependence of the hydrodynamic mode
amplitude $x(t)$ is determined by the Langevin equation
\begin{equation}
\dot x = f_0(x) + g_0(x)\/ \zeta (t).\label{1}
\end{equation}
Here the dot indicates the differentiation with respect to time $t$,
deterministic force $f_0(x)$ is defined according to the thermodynamic
potential $V_0(x)$ as $f_0=-\partial V_0(x)/\partial x$, $\zeta(t)$ is the
colored multiplicative noise with amplitude defined by function $g_0(x)$. A
specific form for the noise $\zeta(t)$ we choose in kind of Ornstein--Ulenbeck
process, i.e.
\begin{equation}
\tau\dot\zeta= -\zeta + \xi(t),\label{2}
\end{equation}
where $\tau$ is a self--correlation time, $\zeta$ is the Gaussian distributed
variable with zero mean and exponentially decaying correlations:
\begin{equation}
\langle\zeta(t)\zeta(t')\rangle=(1/2\tau)\exp(-|t-t'|/\tau),\label{3}
\end{equation}
$\xi(t)$ is a white noise---namely, Gaussian stochastic variable with zero mean
and $\delta$--correlated: $\langle \xi(t) \xi(t')\rangle =\delta(t-t')$. The
amplitude $g_0(x)$ of the multiplicative noise in Eq.(\ref{1}) is chosen in the
simple form
\begin{equation} g(x)=|x|^a,\label{4}
\end{equation}
where $a$ is a positive exponent varying from 0 to 1 \cite{UFN}. We shall
describe our system by master parameter, self--correlation time and velocity of
noise growth defined by exponent $a$.

If we take time derivative of Eq.(\ref{1}), replace first $\dot\zeta$ in terms
of $\zeta$ and $\xi$ from Eq.(\ref{2}) and then $\zeta$ in terms of $\dot x$
and $x$ from Eq.(\ref{1}), we obtain the following non--Markovian stochastic
differential equation (SDE) (see Ref.\cite{cond-mat}):
\begin{equation}
\tau\left(\ddot x-\dot x^2\partial_xg_0(x)/g_0(x)\right)=-\sigma(x)\dot
x+f_0(x) + g_0(x)\/ \xi(t),\label{5}
\end{equation}
where
\begin{equation}
\sigma(x)=1-\tau f_0(x)\partial_x\ln\left[{f_0(x)\over g_0(x)}\right]
\label{5a}
\end{equation}
(throughout the paper, the It\^{o} interpretation for the SDE will be meant).
In the limit $\dot x^2=0$ a system behavior was described in
\cite{Shapiro,UFJ}. Here, following ``unified colored noise approximation''
Ref.\cite{UCNA} we can recover a Markovian SDE. It needs to use adiabatic
elimination (neglecting $\ddot x$) and to neglect $\dot x^2$ so that the
system's dynamics be governed by a Fokker--Planck equation. In order to make
correct transition to an ordinary, linear in $\dot x$, SDE we have to use
It\^{o}'s differential rule \cite{VanKampen}. In above mentioned suppositions
SDE (5) takes form
\begin{equation}
\sigma(x) \dot x=f_0(x)+g_0(x)\xi(t).\label{6a}
\end{equation}
Expression in lhs in Eq.(\ref{6a}) defines time derivative for variable $z$.
The relation between $z$, and $x$ is set by expression ${\rm d}z=\sigma(x){\rm
d}x$. An equivalent equation to Eq.(\ref{6a}) can be write down as
\begin{equation}
\dot x = f(x)+ h(x)+ g(x)\xi(t),\label{7_1}
\end{equation}
with
\begin{equation}
f(x)=\frac{f_0(x)}{\sigma(x)},\qquad%
h(x)=-{1\over
2}\left(\frac{g_0(x)}{\sigma(x)}\right)^2\partial_x\ln\sigma(x),\qquad
g(x)={g_0(x)\over \sigma(x)}.\label{7_2}
\end{equation}
Here we stress that drift term $h(x)$ appears only by accounting the noise
self--correlation time.

Let us now go over to a new field $y(t)$ connected with the initial field
$x(t)$ through the relation
\begin{equation}
{{\rm d}x\over {\rm d}y}=g(x(y)).\label{6}
\end{equation}

The new stochastic field taking into account Eq.(\ref{6}) satisfies the
Langevin equation
\begin{equation} \dot y = \tilde h(x(y))+\xi,\label{8}
\end{equation} with additive noise and effective force
\begin{equation}
\tilde h\equiv (f+h)/g-\partial_xg/2.\label{9} \end{equation}

The obtained equation (\ref{8}) allows us to use the standard field scheme
\cite{ZJ} based on analysis of the generating functional. The latter has the
form of the functional Laplace transform
\begin{equation}
Z\left\{u(t) \right\}=\int Z\left\{y(t) \right\}\exp\left(\int u y {\rm d}t
\right)\mathcal{D}y (t)\label{10} \end{equation} for the partition function
\begin{equation} Z\left\{y(t) \right\}=\left \langle \prod_{t}\delta \left\{ \dot y
 -\tilde h -\xi \right\}\det \left| {{\delta \xi}\over {\delta y}}
\right| \right \rangle_{\xi},\label{11} \end{equation} where $\mathcal{D}y$
denotes integration over all paths starting at $y(0)$ for $t=0$ and ending at
$y(t_f)$ for $t=t_f$. It is defined as
\begin{equation}
\mathcal{D}y=\lim\limits_{N\to\infty}\prod\limits_{i=1}^{N-1}{\rm d}y(t_i),
\end{equation}
where $y(t_i)$ is a field at time $t_i=i\epsilon$, having sliced the interval 0
to $t_f$ in $N$ parts of size $\epsilon=t_f/N$. The argument of the
$\delta$-function in Eq.(\ref{11}) can be reduced to the Langevin equation
(\ref{8}), and the determinant ensuring a transition from continual integration
with respect to $\xi(t)$ to $y (t)$ is equal to unity in It\^{o} calculus.

In the approach developed by Zinn-Justin \cite{ZJ}, $n$--fold variation of
functional (\ref{10}) over the auxiliary field $u(t)$ makes it possible to find
the $n$-th order correlator for the hydrodynamic mode amplitude $y(t)$ and to
construct the perturbation theory. We shall proceed, however, from expression
(\ref{11}) for the conjugate functional $Z\{y (t) \}$ whose variation leads to
the most probable realization of the stochastic field $y(t)$.

Going over to an analysis of functional (\ref{11}), we write the
$\delta$-function in integral form \begin{equation} \delta \left\{
y(t)\right\}=\int \limits_{-i\infty}^{i\infty}\exp\left(-\int q y{\rm d}t
\right)\mathcal{D}q.\label{12}
\end{equation} Averaging over the noise $\xi$ with the help of the Gaussian
distribution
\begin{equation}
P_0\{\xi\} \propto \exp \left\{-{1\over{2}} \int \xi^2 (t) {\rm d}t
\right\},\label{13} \end{equation} corresponding to (\ref{3}) and taking into
account Eq.(\ref{12}), we reduce functional Eq.(\ref{11}) to the standard form
\begin{equation}
Z\left\{y(t) \right\}=\int P\{y(t), q(t) \} \mathcal{D}q, \quad P\equiv
e^{-\mathcal{S}}. \label{14}
\end{equation}
 Here
the probability distribution $P\{y,q\}$ is specified by the action
$\mathcal{S}=\int \mathcal{L} {\rm d}t$, where the Lagrangian is given by
\begin{equation}
\mathcal{L}(y,q)=q \left(\dot y -\tilde h \right) -q^2/2. \label{15}
\end{equation}

We shall use in subsequent analysis the Euler equations
\begin{equation}
 {{\partial \mathcal{L}}\over{\partial z}} -{{\rm d}\over{{\rm d}t}}{{\partial
\mathcal{L}}\over{\partial \dot z}}= {{\partial \mathcal{R}}\over{\partial \dot
z}},\quad z\equiv\{y,q\}, \label{16} \end{equation} in which the dissipative
function is defined as \begin{equation} \mathcal{R}(y)=\dot y^2/2.\label{17}
\end{equation} As a result, the equations for the most probable
realizations of the stochastic fields $y(t)$ and $q(t)$ assume the form
\begin{eqnarray}
&&\dot y = \tilde h + q,\label{18}\\ &&\dot q = - q\left(1 + \partial_x \tilde
h \right)-\tilde h. \label{19}
\end{eqnarray}

A comparison of Eq.(\ref{18}) with the stochastic equation (\ref{8}) having the
same form shows that the fields $y(t)$ and $q(t)$ are the most probable values
of amplitudes of the auxiliary hydrodynamic mode defined by relation (\ref{6})
as well as by fluctuation of the conjugate force. Obviously, the latter can be
reduced to the conjugate momentum $q=\partial \mathcal{L}/\partial \dot y$.

In order to return to the initial stochastic field $x(t)$, we shall use the
relation (\ref{6}) and definition (\ref{9}). As a result, the Lagrangian
(\ref{15}) and the dissipative function (\ref{17}) assume the form
\begin{equation} \mathcal{L}(x,\phi)=\phi \left(\dot x - f-h
+ g\partial_x g/2\right) -g^2\phi^2/2, \label{20} \end{equation}
\begin{equation}
\mathcal{R}(x)=\dot x^2/2g^2,\label{21}
\end{equation} where we have used the definition of the conjugate momentum
$\phi=\partial \mathcal{L}/\partial \dot x$ leading to the relation
\begin{equation} \phi=q/g.\label{22}
\end{equation} In this case, the Euler
equations (\ref{16}) assume the form
\begin{equation}
\dot x = f + h - g\partial_x g/2+ g^2\phi,\label{23}
\end{equation}
\begin{equation}
\dot \phi = -\phi\Bigl[1+\partial_x f+\partial_x h-\partial_x(g\partial_xg)/2
+\phi g\partial_x g\Bigr]-(f+h)/g^2+\partial_x g/2g.\label{24}
\end{equation} Similar equations can be obtained directly from the system
(\ref{18}), (\ref{19}) using the relations (\ref{6}) and (\ref{22}) as well as
definition (\ref{9}).

In order to calculate noise self--correlation time influence on the system
behavior we shall use the $x^4$-model for potential
\begin{equation} V_0(x)={\varepsilon \over 2} x^2 +{1\over 4}x^4,\qquad
\varepsilon\in[-1,~1] \label{26} \end {equation} and the definition (\ref{4})
of multiplicative function. Since the Lagrangian (\ref{20}) does not change its
form upon simultaneous reversal of the signs of $x$ and $\phi$, the phase
portraits will possess central symmetry relative to the origin $x = \phi = 0$.
On the other hand, the axis $x = 0$ on which the noise intensity assumes zero
value is singular in accordance with (\ref{4}). For this reason, we can confine
our analysis only to the upper part of the phase plane corresponding to the
value of $x > 0$.

First of all let us investigate steady states of the system. In the stationary
case $\dot x = \dot \phi = 0$, we have two equations
\begin{equation}
\phi=-\frac{\sigma f_0}{g_0^2}+{1\over 2}\partial_x\ln g_0,\label{29}
\end{equation}
\begin{equation}\label{30}
\phi\left\{%
\frac{\sigma f_0}{g_0^2}\partial_x\ln\left[{f_0\over g_0}\right]+{1\over 2
}(\partial_x\ln\sigma)\partial_x\ln g_0-\frac{\partial_x^2g_0}{2g_0}\right\}=0.
\end{equation}
From Eq.(\ref{29}) it follows that at noise self--correlation time not
exceeding the value $\tau_0$ (Fig.1) the form of the phase portrait is
characterized by the presence of a single saddle point $S$ whose position is
defined by the solution of Eq.(\ref{30}) and corresponds to $\phi\ne 0$.
Figure~2 shows that bifurcation takes place at $\phi=0$ at noise
self--correlation time $\tau_0$ (Fig.2a, 2b). This is accompanied by the
emergence of an additional saddle point $S$ and an attractive node $C$ whose
positions are determined by the condition $\phi = 0$ and specified by the
coordinates $x_\mp$.

In Fig.3 we show critical magnitude of the noise self--correlation time
$\tau_c$ vs temperature $\varepsilon$ (as maximum of the dependence
$\tau_0(a)$). As shown in Fig.1, above $\tau_c$ the system is always ordered
for any velocity of the noise growth, defined by exponent $a$. Obviously, the
saddle point $S$ and the node $C$ merge at the point corresponding to the
correlation time $\tau_c$.

The above analysis shows that in systems with $\tau <\tau_0$ the stationary
state corresponds to the point $S$ whose coordinates are defined as
\begin{equation}
\phi=-\frac{f+h}{g^2}+{1\over 2}\partial_x g,\label{35a}
\end{equation}
\begin{equation}
\partial_x(f+h)=-\phi g\partial_x g+{1\over 2}\partial_x[g\partial_x g]
.\label{35b}
\end{equation}
The meaning of these coordinates becomes clear if we proceed to the additive
limit $g(x)\to 1$: condition (\ref{35a}) indicates that the most probable value
$\phi$ of the fluctuation amplitude of conjugate force for homogeneous systems
has the sign opposite to that of $f+h$; according to Eq.(\ref{35b}), the
``susceptibility'' $(\partial_x^2\widetilde V)^{-1}=-1/\partial_x(f+h)$ in this
case assumes an infinitely large value. Thus, node $S$ corresponds to the
stationary state of a thermodynamic system which is unstable to a transition to
the ordered phase. Then the phase portrait has the form shown in Fig.4a.

At temperatures $\tau >\tau_0$, the steady state with the coordinates
\begin{equation}
\phi=0,\label{36a}
\end{equation}
\begin{equation}
2\sigma f_0=g_0\partial_x g_0.\label{36b}
\end{equation}
is formed at the point $C$. In the additive limit, this point corresponds to
the state of thermodynamic equilibrium. Figure 4 shows that the corresponding
phase portrait can be obtained not only upon an decrease in temperature (see
Fig.4a). Here we stress that the system can pass from equilibrium (ordered)
domain trough unstable (disordered) to stable if we increase the velocity of
the noise growth (see Fig.1 and Fig.4b).

Naturally, a real thermodynamic system in the process of its evolution tends to
the equilibrium state Eqs.(\ref{36a}, \ref{36b}) rather than to the unstable
state Eqs.(\ref{35a}, \ref{35b}). The equilibrium state corresponds to small
and large values of the exponent $a$ for any values of the parameter $\tau$.
The domain of medium values of the exponent $a$ at small noise correlations
corresponds to unstable states.

Comparing Fig.4a, Fig.4b we see that they are distinguished by the location of
the attraction node on the axis $\phi = 0$. The general form of phase portraits
shown in Fig.4 is characterized by the presence of two separatrices with
branches $PQ$ and $MS_0N$. They divide the phase plane into three isolated
regions corresponding to large, intermediate, and small values of the order
parameter $x$. The first region is characterized by an indefinite increase in
the values of the quantities $x$ and $\phi$ with time $t\to \infty$. It will be
sown below that this is not realized in actual practice. The region of
intermediate values of $x$ in which the system ultimately goes over to a
stationary ordered state is most interesting. It is this region that determines
the phase transition kinetics. The formation of the region corresponding to
values $x\ll 1$ is associated with the multiplicative nature of noise. In this
region, the order parameter $x(t)$ tends with time to the value $x = 0$.

Let us analyze the behavior of the system in each of these regions. For this
purpose, we consider the probability of realization of a phase trajectory
corresponding to different initial values $x_0\equiv x(t=0)$. In accordance
with (\ref{14}), the probability can be written in the form
\begin{equation} P(x_0)\propto \exp\left\{-{1\over
2}\int\limits_{x_0}g^2\phi^2 {\rm d}t\right\},\label{37} \end{equation} where
expressions (\ref{20}) and (\ref{23}) are taken into account and integration is
carried out along the corresponding trajectory. The dependence $P(x_0)$
obtained for an exponent $a < 1/2$ is shown in Fig.5 (curves 1, 2). Apart from
the trivial increase in probability (\ref{37}) which approaches the origin for
values of $x_0$ corresponding to separatrices, the jumps near which the value
of $P(x_0)$ can increase insignificantly are observed. Outside the region
bounded by the (outer) separatrix, we have $P = 0$ since $x(t)$, $\phi(t)\to
\pm\infty$ for $t\to \infty$ in this case.

Such a behavior of the probability $P(x_0)$ can be explained by the form of the
time dependencies $x(t)$ and $\phi(t)$ during relaxation of the initial value.
Far away from the region of $S_0$ (see Fig.4a), the quantities $\phi$ and $x$
rapidly change their values, the change slowing down as we approach this
region. Such a behavior can be explained by the fact that the action
$\mathcal{S}\{ x(t), \phi(t)\}$ changes much more slowly near this region than
away from it. This can be visualized by associating the region $S_0CS$ with the
bed of a large river \cite{10}.

For values of the exponent $a > 1/2$, the integrand in formula (\ref{37})
diverges, and the probability $P$ assumes zero value for $x_0\ll 1$. Typically,
this divergence is observed only in the region of phase portrait bounded by the
separatrix branch $S_0O$ in Fig.4b.

In order to explain the form of the dependence $P(x_0)$, we analyze the
behavior of the quantities $x(t)$ and $\phi(t)$ for various values of exponent
$a$. For this purpose, we put $\dot \phi = 0$ in Eq.(\ref{24}). The obtained
quadratic equation gives stationary values of conjugate momentum in the limit
$x\to 0$:
\begin{equation}\label{38} \phi=\left\{\begin{array}{ll}{1\over 2}\left({1\over
2}-a\right)^{-1}x^{1-2a},& a<{1\over 2};\\ \left(a-{1\over 2}\right)x^{-1}, &
a>{1\over 2}.\end{array}\right.
\end{equation}

 Thus, for $a < 1/2$, the system tends with time to the origin
$x = \phi = 0$, and the attraction node jumps to infinity ($\phi\to\infty$,
$x=0$) as $a$ exceeds the critical value $a = 1/2$. The corresponding integrand
in distribution (\ref{37}), i.e.,
\begin{equation} \label{39}
g^2\phi^2=\left\{\begin{array}{ll}{1\over 4}\left({1\over
2}-a\right)^{-2}x^{2(1-a)},& a<{1\over 2};\\ \left(a-{1\over
2}\right)^{2}x^{-2(1-a)},& a>{1\over 2}\end{array}\right.
\end{equation}
is characterized by the sign inversion in the exponent upon a transition
through the critical value $a = 1/2$. Substituting Eq.(\ref{38}) for $a < 1/2$
into the equation of motion (\ref{23}) and retaining the leading term in it, we
obtain
 the equation $2\dot x=-ax^{2a-1}$ which gives the time dependence
of the order parameter: \begin{equation} x^{2(1-a)}=a(1-a)(t_0-t),\qquad
t<t_0,\qquad a<1/2,\label{40}
\end{equation} where $t_0$ is the integration constant defining the
time during which the point gets to the axis $x = 0$. The substitution of
dependence Eq.(\ref{40}) into Eq.(\ref{39}) at $a < 1/2$ and of the obtained
expression into the integral in Eq.(\ref{37}) shows that the probability
$P(x_0)$ of realization of a trajectory in the region $x\ll 1$ of the phase
portrait differs from zero (see Fig.5, curves 1, 2).

A completely different situation is observed for an exponent $a > 1/2$. In this
case, Eq.(\ref{23}) can be reduced to $2\dot x=-(1-a)x^{2a-1}$, leading to
Eq.(\ref{40}) in which the factor $a(1-a)$ is replaced by $(1-a)^2$. However,
for $a
> 1/2$ the expression $g^2\phi^2$ acquires an exponent with the opposite sign in
accordance with Eq.(\ref{39}) so that it assumes an infinitely large value for
$t\to t_0$. As a result, the probability Eq.(\ref{37}) becomes vanishingly
small (see Fig.5, curves 3, 4). The physical reason behind such a behavior is
that the system gets to the axis $x = 0$ over a finite time interval
$t_0<\infty$, which ensures an infinitely large value of the conjugate momentum
$\phi\propto x^{-1}\propto (t_0-t)^{-1/2(1-a)}$. This can be visualized as the
precipitation of condensate (absorbing state) of configuration points from the
phase portrait domain $x\ll 1$ onto the abscissa axis for $\phi\to\infty$. Note
that the condition $t_0<\infty$ is satisfied only below the separatrix branch
$S_0O$ in Fig.4b, while in the region $NS_0O$ we have $t_0=\infty$, and the
convergence of the integrand in (\ref{37}) is not manifested. Consequently, the
equality $P = 0$ holds only below the curve $S_0O$ (see Fig.5 and Fig.4b).

In accordance with (\ref{40}), the time dependence of the most probable
magnitude $x(t)\propto t^H$ is defined by the exponent $H^{-1}=2(1-a)$ whose
magnitude determines the fractional dimension $D\equiv H^{-1}$ characterizing
the domain $x\ll 1 $ of the phase portrait of a system with multiplicative
noise \cite{11}. In the additive limit $a = 0$, we have dimension $D = 2$ of
the phase plane as expected. This means that as the time $t\to\infty$, the
phase trajectories of the system fill the entire phase plane. The increase of
the exponent $a > 0$ leads to a decrease in the fractional dimension $D$ which
assumes the critical value $D = 1$ for $a = 1/2$. As the value of $a$ increases
further, the fractional dimension of the set of points on the plane $x, t$,
which is the law of motion $x(t)$, becomes smaller than unity. The physical
reason behind such a behavior is the above mentioned absorption of
configuration points by the axis $x = 0$ for $\phi\to\infty$.

Finally, it must be stressed that the noise correlation time can play role of
master parameter as the temperature and produces transition to the ordered
state. Moreover we shown that the increasing of the velocity of the noise
intensity growth produces appearance of the domain of disordered states. This
domain disappears if noise self-correlation time does not exceed value
$\tau_c$. Path integral solution shown that absorbing state appearance
 is characterized by the vanishingly small probability of the system states realization.
Asymptotic time dependence of the order parameter $x$ explains picture of
absorbing states by the fractal dimension smaller than unity.

\section*{Acknowledgement}
I am grateful to prof.~A.I.Olemskoi for inspiring discussions and helpful
comments.

\newpage
\centerline{\bf FIGURE CAPTIONS} \vspace{1cm}

{\bf Fig.1.} Phase diagram indicating appearance of the ordered phase ($x\ne
0$): noise correlation time $\tau$ vs exponent $a$ (curves 1, 2 correspond to
$\varepsilon=0.65$, 0.7).

{\bf Fig.2.} Bifurcation diagrams: stationary values of the order parameter vs
noise self-correlation time at $\varepsilon=0.7$ ($a=0.2$ (a); $a=0.6$, (b)).

{\bf Fig.3.} Phase diagram indicating the ranges of the parameters $\tau_c$,
$\varepsilon$.

{\bf Fig.4.} Basic types of phase portraits in the ordered state:
$\varepsilon=0.7$, $\tau = 0.4$ ($a = 0.2$ (a), $a = 0.8$ (b)). The notation
for stationary points is the same as in Fig.2.

{\bf Fig.5.} Dependence of the probability $P$ of realization of various
trajectories on the initial value of the order parameter $x_0$ for $\varepsilon
= 0.7$ (curves 1, 2 correspond to $a = 0.2$ $\tau = 0.4$, 0.6, curves 3,4
correspond to $a = 0.8$ $\tau = 0.4$, 0.6. The initial value of the conjugate
momentum $\phi_0 = 0.5$.

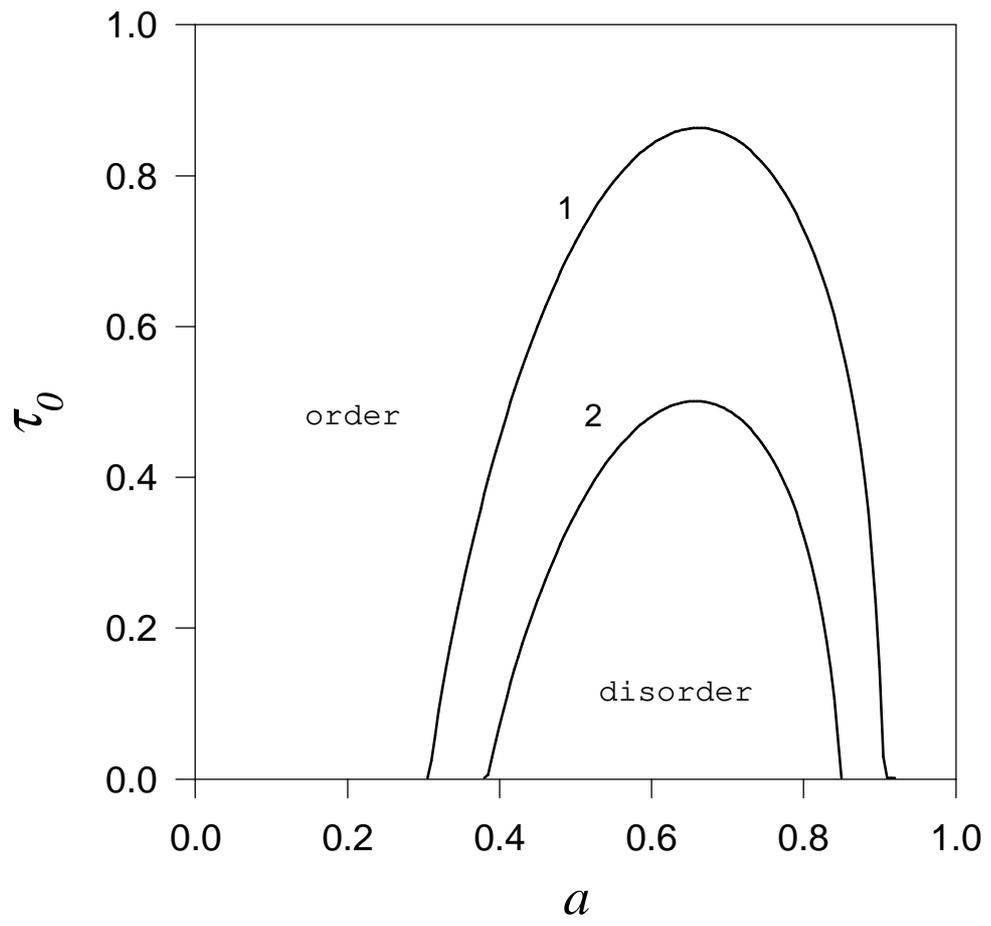
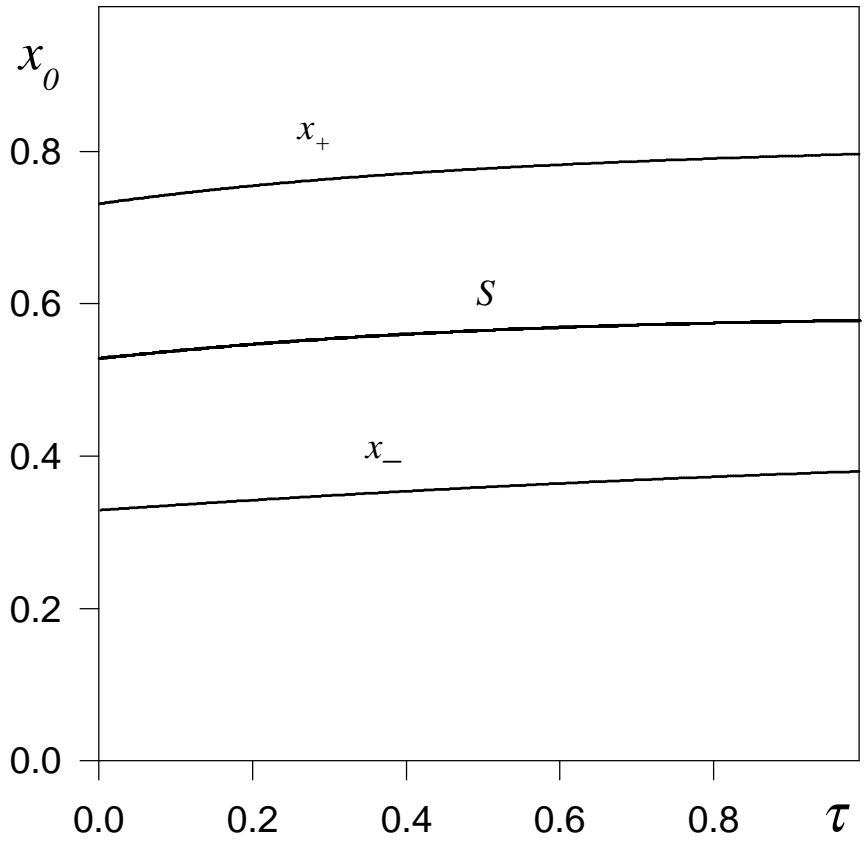
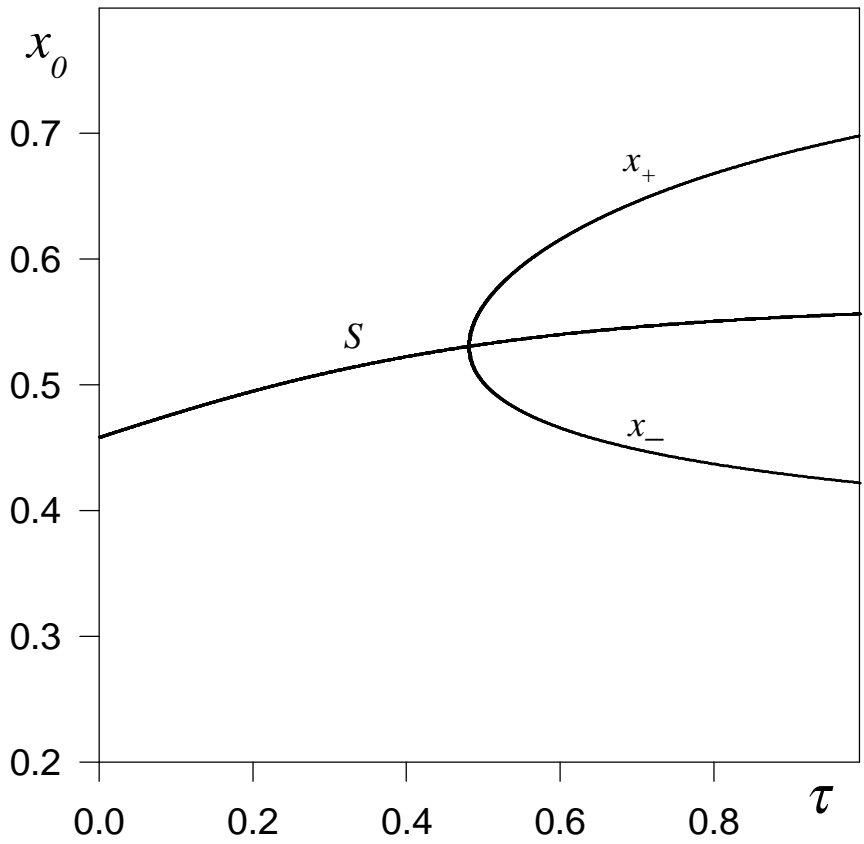
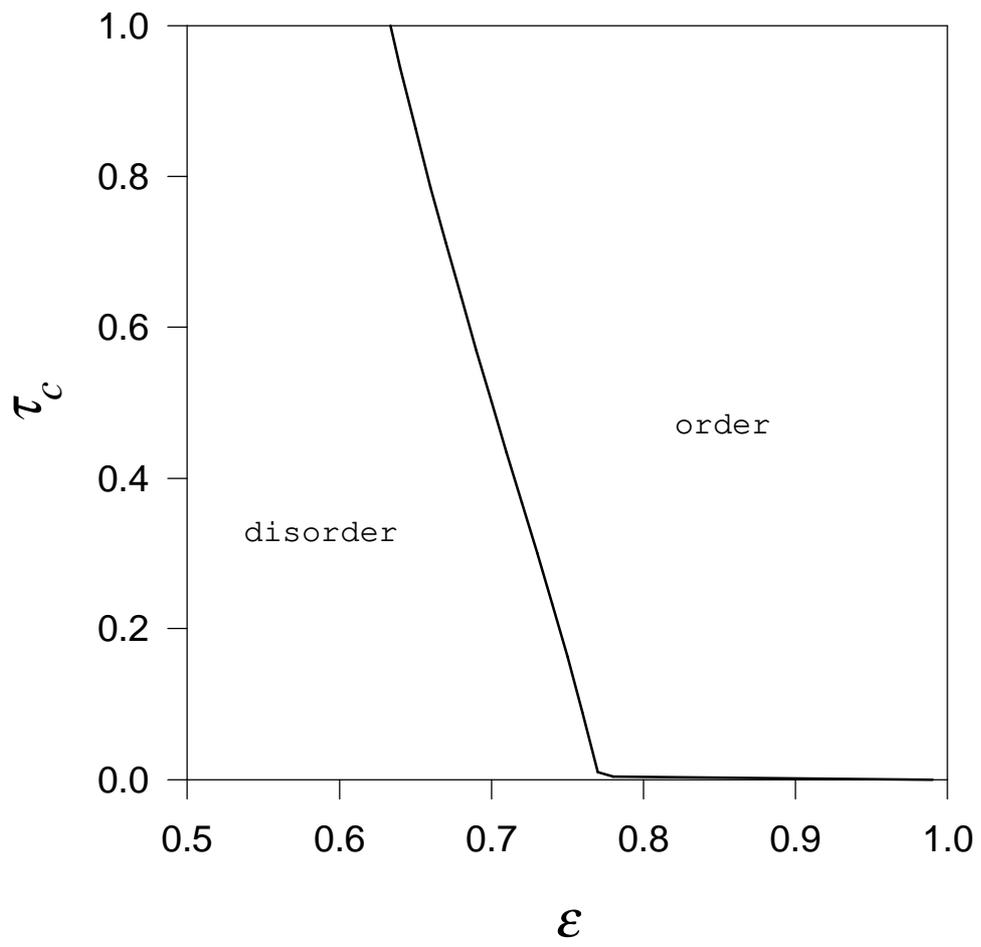
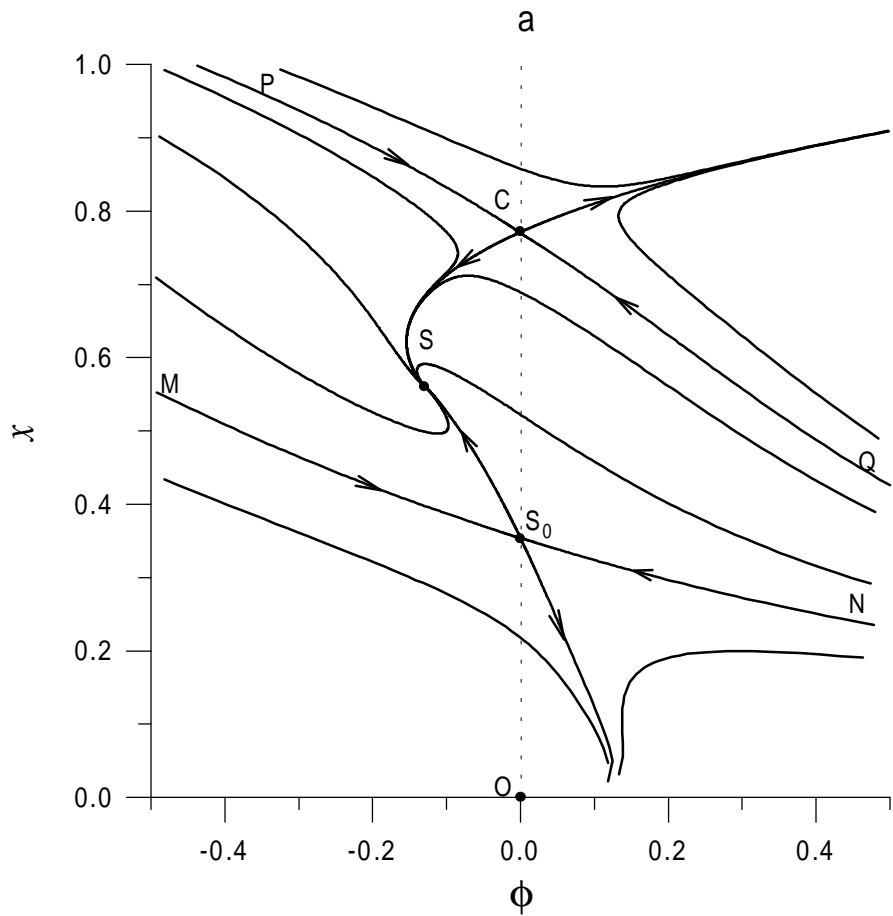
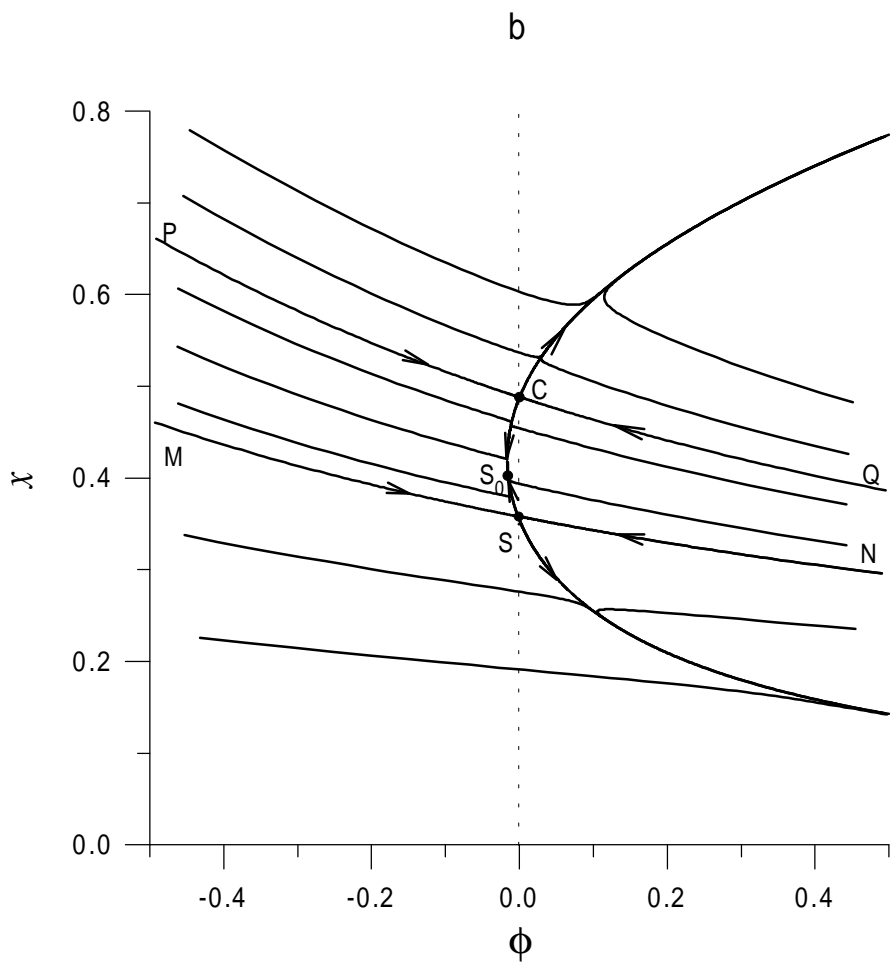
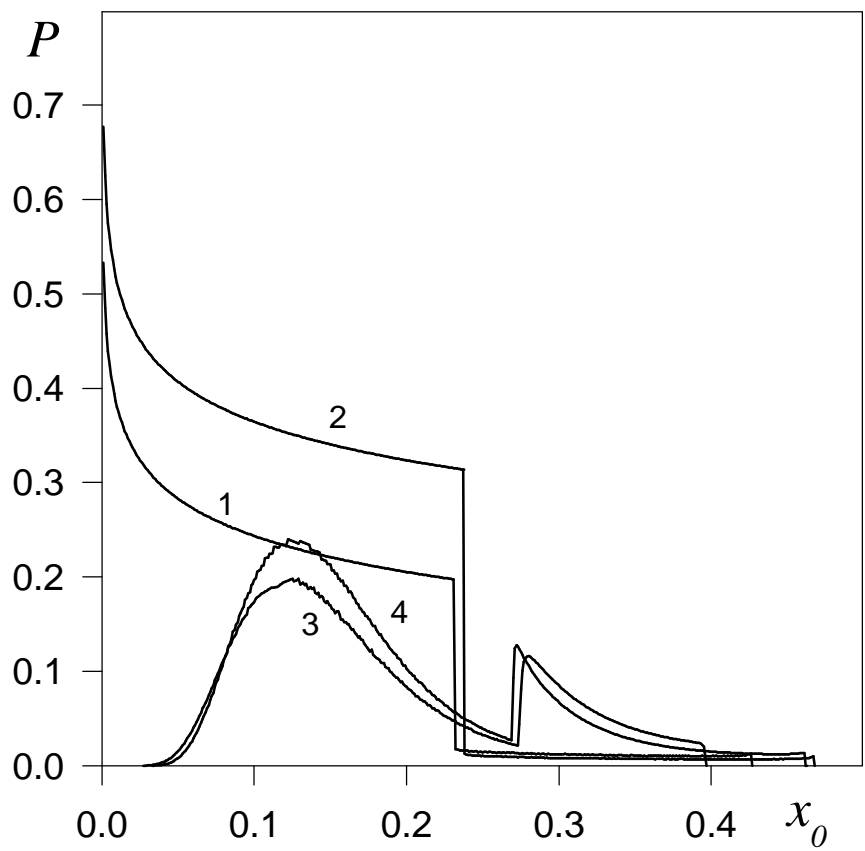
\end{document}